\documentclass{WileyMSP-template}

\begin{document}

\title{Evaporation characteristics of Er$^{3+}$ doped silica fiber and its application in the preparation of whispering gallery mode lasers}

\maketitle

% Author: Please give full first and last names for authors and include * after the name of all corresponding authors

\author{Angzhen Li},
\author{Jonathan M. Ward},
\author{Ke Tian},
\author{Jibo Yu},
\author{Shengfei She},
\author{Chaoqi Hou},
\author{Haitao Guo},
\author{S\'{i}le {Nic Chormaic}*},
\author{Pengfei Wang*}

% Dedication

% Affiliations: Please provide adacemic titles (Prof. or Dr.) for all authors where applicable, and include an institutional email address for all corresponding authors
\begin{affiliations}
Angzhen Li\\
Tianjin Key Laboratory of Quantum Optics and Intelligent Photonics, School of Science, Tianjin University of Technology, Tianjin 300384, China\\

Jonathan M. Ward\\
Physics Department, University College Cork, Cork, Ireland\\

Ke Tian, Pengfei Wang\\
Key Laboratory of In-Fiber Integrated Optics of Ministry of Education, Harbin Engineering University, Harbin 150001, China\\

Email Address: pengfei.wang@tudublin.ie

Jibo Yu\\
Xi’an Institute of Applied Optics, Xi’an 710065, China\\

Shengfei She, Chaoqi Hou, Haitao Guo\\
State Key Laboratory of Transient Optics and Photonics, Xi’an Institute of Optics and Precision Mechanics, Chinese Academy of Sciences, Xi’an 710119, China\\

S\'{i}le {Nic Chormaic}\\
Okinawa Institute of Science and Technology Graduate University, Onna, Okinawa 904-0495, Japan\\

Email Address: sile.nicchormaic@oist.jp

\end{affiliations}

% Keywords: Please provide a minimum of three and a maximum of seven keywords, separated by commas

\keywords{whispering gallery modes, microsphere resonators, silica, rare earth ions}

% Abstract should be written in the present tense and impersonal style (i.e., avoid we), and be at most 200 words long
\begin{abstract}

The fabrication of whispering gallery lasers (WGL) is used to experimentally evaluate the evaporation rate (mol/$\mu$m) and ratio (mol/mol) of erbium and silica lost from a doped fiber during heating. Fixed lengths of doped silica fiber are spliced to different lengths of undoped fiber and then evaporated by feeding into the focus of a CO$_{2}$ laser. During evaporation, erbium ions are precipitated in the doped silica fiber to control the erbium concentration in the remaining SiO$_2$, which is melted into a microsphere. By increasing the length of the undoped section, a critical point is reached where effectively no ions remain in the glass microsphere. The critical point is found using the lasing spectra of the whispering gallery modes in microspheres with equal sizes. From the critical point, it is estimated that, for a given CO$_{2}$ laser power, $6.36 \times 10^{-21}$~mol of Er$^{3+}$ is lost during the evaporation process for every cubic micron of silica fiber. This is equivalent to $1.74 \times 10^{-7}$~mol of Er$^{3+}$ lost per mol of SiO$_{2}$ evaporated. This result facilitates the control of the doping concentration in WGLs and provides insight into the kinetics of laser-induced evaporation of doped silica. 

\end{abstract}

% Text: Please use section headings and subheadings as specified below. For communications, all section headings apart from Experimental Section should be removed
% Please make the first reference to a display item bold: \textbf{Figure 1}
% Do not abbreviate Figure, Equation, etc.; display items are always singular, i.e., Figure 1 and 2.
% Equations are always singular, i.e., Equation 1 and 2, and should be inserted using the {equation} environment, not as graphics
% Please do not use footnotes in the text, additional information can be added to the Reference list.

\section{Introduction}

Deposition is the opposite of sublimation; however, both of these phenomena are used in tandem to create thin films and nanoparticles on substrates. In particular, the deposition of rare earth ion doped silica (SiO$_{2}$) glass has a wide range of applications in many fields from telecommunications and biosensing to nonlinear and quantum optics. Probably the most well-known and commonly used method is pulsed laser deposition \cite{chrisey1994pulsed, lahoz2008upconversion, morea2016pulsed, doi:10.1063/1.5097506, HASABELDAIM2020144281}. A doped target material is ablated by irradiation with a pulsed laser and the resulting plume emanating from the target is intercepted by a substrate. The particles in the plume deposit on the substrate where they can form a thin film. The doping concentration and thickness of the film can be partially controlled by the laser power and number of pulses \cite{morea2016pulsed, D2CS00938B}.
Inductively coupled plasma-optical emission spectrometry and laser ablation inductively coupled plasma mass spectrometry are methods used to determine the elemental constituent of glass \cite{schenk2012elemental} and the relative concentrations of any dopants. Laser ablation was used to examine historical glass \cite{van2009multi, corzo2018use, doi:10.1021/je200693d, doi:10.1366/000370210790918346, mccormack1971vapor, habermann1964vapor, sysoev2003high, Elhadj:12} as well as in forensic studies \cite{corzo2018use}. These analytical methods use ablation but do not reveal any information about the evaporation kinetics of the individual components.

Vaporisation of glass materials is also a well studied effect with a number of papers presenting vapour pressures and evaporation coefficients for SiO$_2$ and Si mixes \cite{doi:10.1021/je200693d, doi:10.1366/000370210790918346} as well as pure erbium \cite{mccormack1971vapor} and other rare earth ions \cite{habermann1964vapor}. There is a nonlinear relationship between vapor pressure/evaporation and temperature which is traditionally measured using a Knudsen effusion cell and a thermogravimetric balance \cite{mccormack1971vapor, habermann1964vapor}. Continuous wave or pulsed lasers are both used to create evaporation from glass targets for such studies, CO$_2$ lasers are particularly well-suited for heating glass materials, with several studies on evaporation rates reported in the literature \cite{sysoev2003high, Elhadj:12, doi:10.2351/7.0000482}. Due to the properties of silica, the rare earth ion concentration in doped silica materials is usually very low \cite{doi:10.1179/1743280412Y.0000000005, WANG201990}. Therefore, the change of ion concentration due to the evaporation process of doped silica is difficult to  measure simply and directly.  A typical method used is Rutherford backscattering spectrometry after the pulsed laser deposition, to determine substrate Er$^{3+}$ concentration dependence on the number of pulses \cite{serna2005improving}. Surprisingly, there seems to be little information on the evaporation of rare earth ion doped glass, particularly the evaporation ratio of the dopant to the glass host. To the best of our knowledge, the characteristics of SiO$_2$ doped with trace rare earth ions during evaporation have not been studied. Therefore, we present a measurement of the evaporation rate of Er$^{3+}$ to SiO$_2$ in a doped glass using a novel method based on whispering gallery spectroscopy and volumetric evaporation analysis that does not require any knowledge of the temperature.
 
In this paper, we use whispering gallery resonators (WGRs) as a tool to determine the number of Er$^{3+}$ lost for every cubic micron of SiO$_2$ evaporated. The combination of ultrahigh quality (Q-) factor and small mode volume in WGRs significantly enhances the light-matter interactions, giving them excellent detection/sensing capabilities \cite{2011Detecting, Ward:18, https://doi.org/10.1002/adom.202100143}. To determine the evaporation characteristics, the doped fibers were spliced to sections of undoped fiber, then melted and evaporated to form microspheres. Using different ratios of doped and undoped fiber lengths provides us with a way to gauge the volumes evaporated and to reach a critical doping concentration during the tests. The results show that, for a given CO$_2$ laser power, it is possible to control the doping concentration (by a factor of 20) in the microsphere by adjusting the volumes of fiber evaporated. We also discuss the physical mechanism behind the observed evaporation ratio. 

As an application to our findings, a method to deterministically prepare active microsphere resonators using commercially available doped fiber is presented. Currently, sol-gel coatings are widely used to add rare earth ions to WGRs, but this method needs some careful chemical preparation and the doping concentration control requires repeated sol preparation \cite{Yang:03,doi:10.1063/1.1598623, doi:10.1021/acsphotonics.8b00838}. Other methods for preparing doped WGRs have also been reported, such as doping by ion implantation \cite{KALKMAN2006182} or etching the cladding of commercial doped fiber \cite{8844086}. These methods have not been widely used because of their complex process and low efficiency. Although the preparation of passive silica microspheres by directly melting the end of undoped fiber is widely used \cite{2002Ultralow}, similar methods are not suitable for the preparation of active microspheres by doped fiber. Since the rare earth ions of the doped fiber are only deposited in the core, the microspheres prepared directly from doped fiber often cannot meet the requirements for lasing due to the low doping concentration in the microspheres. In the method proposed in this paper, the different evaporation rates of Er$^{3+}$ and SiO$_{2}$ during heating effectively increase the concentration of Er$^{3+}$ in the microspheres. This boil down or reduction method is not only simple and economical, but can also be used to control and calculate the doping concentration and size of the resulting microspheres. It ensures that any volume of doped optical fiber can be made to produce a microspherical WGR with sufficient concentration of gain material to achieve lasing.

\section{Evaporation characteristics of doped silica}

\begin{figure}[htbp]
\centering\includegraphics[width=18cm]{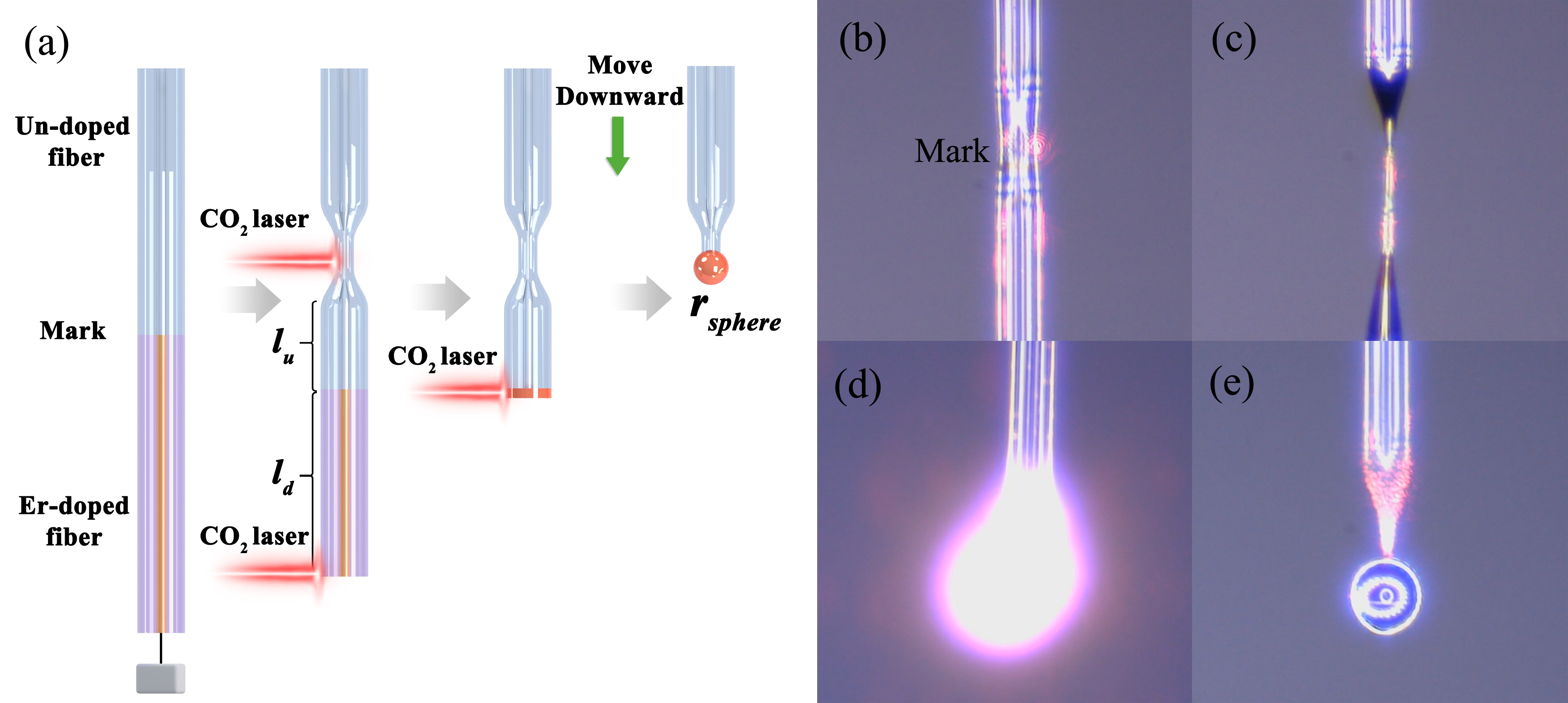}
\caption{(a) Sample preparation flow chart. \emph{l${_u}$} is the length of the undoped fiber section. \emph{l${_d}$} is the length of the erbium doped-fiber section. (b) The fused point is marked by the micro tapering process. (c) A thin rod prepared by SiO$_{2}$ heating. (d) Fiber being heated by high-power CO$_{2}$ laser ($\sim$7 W). (e) Image of a microsphere prepared by evaporation of fused fiber}
\label{fabricate}
\end{figure}

In order to study the evaporation characteristics of doped silica, Er$^{3+}$-doped (custom made) and undoped fibers (SMF-28, Thorlabs) were fused together and then prepared into microspheres. The Er$^{3+}$-doped fiber used in the experiment was prepared through a low-temperature chelate gas phase deposition technique~\cite{9178423}. The microsphere sample preparation process is shown in Figure \ref{fabricate}(a). The fusion splicer was set to continue a slight tapering operation after the splicing process to mark the splicing position between the doped and undoped fiber sections. Figure \ref{fabricate}(b) is a photo of the sample under a microscope, the center of the depression is the splicing point of the two fibers. Next, the fused fiber was placed vertically and a weight was hung on to the end. The undoped fiber was irradiated in the focus of the CO$_{2}$ laser, at a distance \emph{l${_u}$} above the splice mark, until the glass softened and stretched to form a thin rod. Figure \ref{fabricate}(c) shows a photo of the thin rod. Then, at high-power, the CO$_{2}$ laser irradiated the doped fiber at a distance, \emph{l${_d}$}, below the splice mark and cut off the material below. By moving the fiber downward, the tail of the fiber was continuously evaporated by the high-power CO$_{2}$ laser ($\sim$7 W). The laser spot size FWHM was measured as approx 700 $\mu$m which gives a power density of 5~kW/cm$^{2}$. The power density was sufficient to raise the temperature of the glass to near the boiling point and produce rapid evaporation \cite{sysoev2003high, Elhadj:12}. Figure \ref{fabricate}(d) is a photo of the fiber under high-power CO$_{2}$ laser irradiation. Laser heating was stopped when their was eseentially no fiber left below the thin region. This process resulted in the formation of a microsphere connected to the end of the thin rod (the inset of Figure \ref{fabricate}(e), the radius of the microsphere is denoted here as \emph{r${_{s}}$}. The molar percentage of Er$^{3+}$ in the doped fiber core (\emph{p${_{mpc}}$}) was about 0.05 mol$\%$ (Calculated from the external doping method, that is, the molar amount of Er$^{3+}$ in the core was 0.05 mol$\%$ of SiO$_{2}$). The core radius (\emph{r${_{c}}$}) was 1.6 $\mu$m and the fiber radius (\emph{r${_{f}}$}) was 62.5 $\mu$m. The fiber radius of the undoped fiber used in the experiment was also 62.5 $\mu$m, because the value is the same as that of the doped fiber, \emph{r${_{f}}$} is used for both fibers.

\begin{figure}[htbp]
\centering\includegraphics[width=18cm]{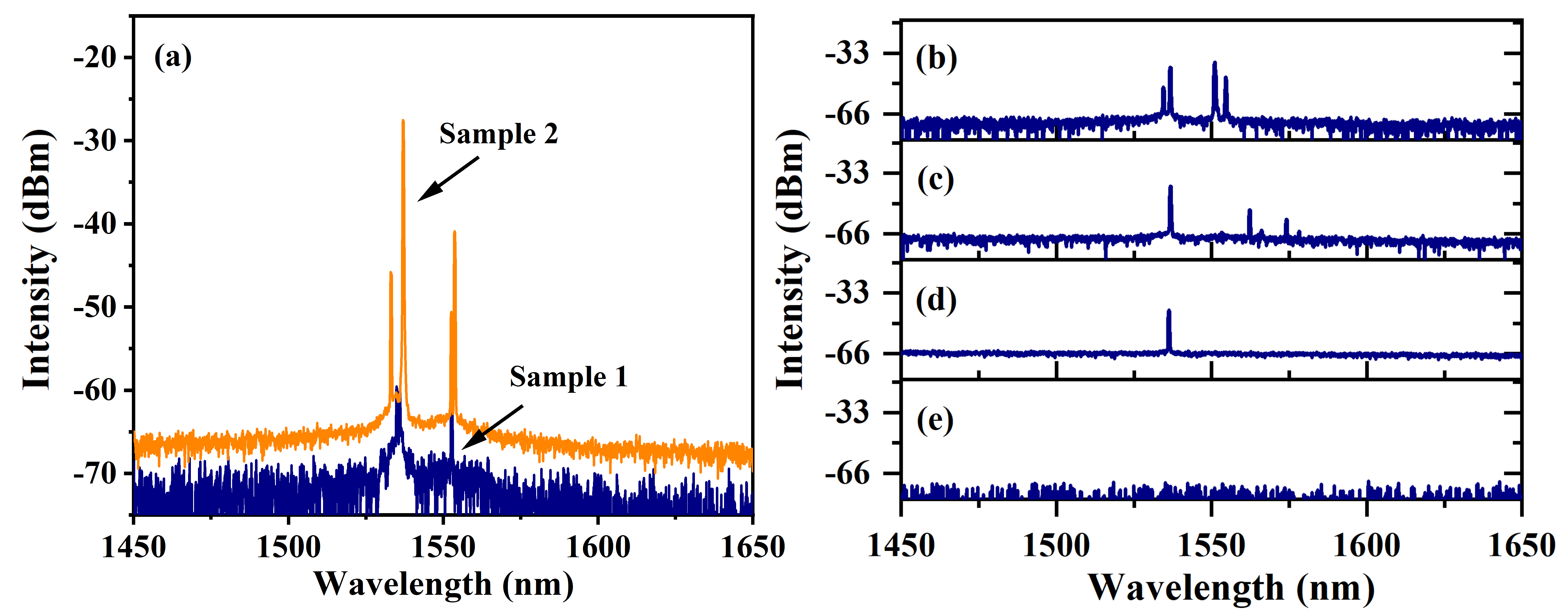}
\caption{(a)Luminescence performance of microspheres prepared by fused fibers with different doped lengths. Parameters of sample 1 are \emph{r${_{s}}$}=71 $\mu$m, \emph{l${_d}$}=1.0 cm, \emph{l${_u}$}=0.4 cm and the parameters of sample 2 are \emph{r${_{s}}$}=70 $\mu$m, \emph{l${_d}$}=3.0 cm, \emph{l${_u}$}=0.4 cm. (b)-(e)Luminescence performance of microspheres prepared by fused fibers with different undoped lengths. Parameters of Sample (b) are \emph{r${_{s}}$}=74 $\mu$m, \emph{l${_d}$}=1.0 cm, \emph{l${_u}$}=0.2 cm; parameters of Sample (c) are \emph{r${_{s}}$}=71 $\mu$m, \emph{l${_d}$}=1.0 cm, \emph{l${_u}$}=0.6 cm; parameters of Sample (d) are \emph{r${_{s}}$}=70 $\mu$m, \emph{l${_d}$}=1.0 cm, \emph{l${_u}$}=0.8 cm; parameters of Sample (e) are \emph{r${_{s}}$}=74 $\mu$m, \emph{l${_d}$}=1.0 cm, \emph{l${_u}$}=0.9 cm.}
\label{lu_and_ld}
\end{figure}

A tapered optical fiber was used for coupling light into the whispering gallery modes (WGMs) of the microsphere \cite{PhysRevLett.85.74}, and was prepared by heating and stretching a single-mode fiber (SMF-28, Thorlabs) to a waist diameter of $\sim$3 $\mu$m. Pump laser light at 980 nm was launched into one end of the tapered fiber and evanescent field coupling to the microsphere was achieved by making contact at the waist of the tapered fiber. The other end of the tapered fiber was connected to an optical spectrum analyzer (OSA, AQ6375, YOKOGAWA). The coupling and test system were the same as described in \cite{Li:19}. Figure \ref{lu_and_ld} shows the spectra of two samples obtained using the same pump power (20 mW) and tapered fiber coupler. The sample parameters are given in Figure \ref{lu_and_ld}(a). Note that the pump power was measured at the input end of the tapered fiber. The sphere sizes and the lengths of the undoped sections, \emph{l${_u}$}, were the same in both samples but the lengths of the doped sections, \emph{l${_d}$}, were different. Whispering gallery lasing (WGL) and fluorescence from the Er$^{3+}$ doped spheres were observed in both samples. 

The bulk of the material forming the microspheres was from the undoped fiber sections and the only source of Er$^{3+}$ in the microspheres was from the doped-fiber sections, which were evaporated. The fact that the microspheres emitted light shows that, during the evaporation of the doped-fiber section, all or part of the Er$^{3+}$ was retained and continuously deposited in the fiber end until it finally mixed into the microsphere. Notably, the microsphere made by evaporating the longer doped-fiber hadstronger luminescence, indicating that evaporation of longer doped-fiber sections lead to more ions being retained. Fixing the length of the undoped section and the sphere size but varying the length of the doped section should allow one to control the concentration of Er$^{3+}$ in the microsphere. 

A second experiment was designed to study whether all or part of the Er$^{3+}$ was retained during evaporation. The same method as described above was used, but this time the length of the doped section was fixed and the length of the undoped section was varied. Figure \ref{lu_and_ld}(b) to Figure \ref{lu_and_ld}(e) show the spectra of four microsphere samples. As seen from these spectra, when the diameters of the spheres were kept the same, the luminescence of the microspheres prepared from the fiber with the longer undoped fiber section was weaker. This shows that Er$^{3+}$ was also lost during the evaporation of the undoped section. The longer undoped sections resulted in lower concentration because a larger volume of glass needed to be evaporated (compared to the shorter undoped section) to achieve the same size sphere. This longer evaporation process caused more Er$^{3+}$ to be lost and a lower concentration in the microsphere.

From this set of spectra, the loss rate of Er$^{3+}$ is estimated. According to Figure \ref{lu_and_ld}(b) to Figure \ref{lu_and_ld}(e), luminescence was no longer detected from the microspheres when the length of the undoped fiber was increased to 0.9 cm (in steps of \emph{l${_d}$}=0.1 cm). Taking into account the ultrahigh Q-factor of the WGM microspheres and the large pump power used in the experiment, it can be inferred that the Er$^{3+}$ concentration of the microsphere in Figure \ref{lu_and_ld}(e) was at an extremely low level. For practical purposes, the Er$^{3+}$ concentration in the microspheres was approximately zero under this condition. The sample in Figure \ref{lu_and_ld}(e) was the dividing point of whether luminescence could be detected or not, which could be regarded as the critical point of complete loss of Er$^{3+}$. An average loss rate, \emph{L${_{mv}}$}, i.e., the number of Er$^{3+}$ lost by evaporation per unit volume from the fiber during the microsphere fabrication can be defined as: 

\begin{equation}
\emph{L${_{mv}}$} =
 \frac{\emph{p${_{mvc}}$}(\pi\times\emph{r${_{c}}$}^2)\emph{l${_d}$}-\emph{p${_{mvs}}$}(4/3)(\pi\times\emph{r${_{s}}$}^3)}{(\pi\times\emph{r${_{f}}$}^2)(\emph{l${_d}$}+\emph{l${_u}$})-(4/3)(\pi\times\emph{r${_{s}}$}^3)}
\end{equation}
Where \emph{p${_{mvc}}$} is the molar volume concentration of Er$^{3+}$ in the doped fiber core, \emph{p${_{mvs}}$} is the molar volume concentration of Er$^{3+}$ in the microspheres. \emph{p${_{mvc}}$} can be calculated from \emph{p${_{mpc}}$}:
\begin{equation}
\emph{p${_{mvc}}$} =
 \frac{\emph{p${_{mpc}}$}[(\pi\times\emph{r${_{c}}$}^2)\emph{l${_d}$}\times\rho]/\rm{M}}{(\pi\times\emph{r${_{c}}$}^2)\emph{l${_d}$}}
\end{equation}
Where $\rho$ is the density of SiO$_{2}$, taken here as $2.2 \times 10^{-12}$ g/$\mu$m$^3$. M is the molar mass of SiO$_{2}$ with a value of 60.084 g/mol. \emph{p${_{mvs}}$} is difficult to measure directly; however, when the Er$^{3+}$ concentration in the microspheres is 0, equation (1) is simplified to:
\begin{equation}
\emph{L${_{mv}}$} =
 \frac{\emph{p${_{mvc}}$}(\pi\times\emph{r${_{c}}$}^2)\emph{l${_d}$}}{(\pi\times\emph{r${_{f}}$}^2)(\emph{l${_d}$}+\emph{l${_u}$})-(4/3)(\pi\times\emph{r${_{s}}$}^3)}
\end{equation}

Equation (3) corresponds to Sample (e) in Figure \ref{lu_and_ld} with \emph{l${_d}$}=1.0 cm and \emph{l${_u}$}=0.9 cm. The corresponding values of the other parameters are: \emph{p${_{mvc}}$}=$1.83 \times$ 10$^{-17}$ mol/$\mu$m$^3$, \emph{r${_{c}}$}= 1.6 $\mu$m, \emph{r${_{s}}$}=74 $\mu$m, \emph{r${_{f}}$}=62.5 $\mu$m. By substituting the above parameters, we get \emph{L${_{mv}}$}=$6.36 \times$ 10$^{-21}$ mol/$\mu$m$^{3}$. This implies that $6.36 \times$ 10$^{-21}$ mol of Er$^{3+}$ was lost in the evaporation process for every cubic micron of fiber.

From another point of view, we can express the ratio of the amount of Er$^{3+}$ to the amount of SiO$_{2}$ evaporated in the preparation process as \emph{R${_{mm}}$} where 
\begin{equation}
\emph{R${_{mm}}$} =
 \frac{\emph{p${_{mvc}}$}(\pi\times\emph{r${_{c}}$}^2)\emph{l${_d}$}}{\rho[(\pi\times\emph{r${_{f}}$}^2)(\emph{l${_d}$}+\emph{l${_u}$})-(4/3)(\pi\times\emph{r${_{s}}$}^3)]/\rm{M}}
\end{equation}

Substituting in the data from Sample (e) in Figure \ref{lu_and_ld}, \emph{R${_{mm}}$}=$1.74 \times$ 10$^{-7}$, this means that $1.74 \times$ 10$^{-7}$ mol of Er$^{3+}$ was lost when each mol of SiO$_{2}$ was evaporated. The experimental phenomena and calculations show that there was a trace loss of Er$^{3+}$ in the high-temperature evaporation process of the Er$^{3+}$-doped SiO$_{2}$ glass. This is an approximate ratio whose accuracy is mainly limited by the step size (0.1 cm) of \emph{l${_d}$} in the experiment, the sensitivity of the spectrum analyzer, and the influence of a small amount of other metal ions in the doped fiber to adjust the refractive index.

\section{A preparation method of microspheres based on evaporation}

As an application of the above research, the method of evaporating Er$^{3+}$ doped fiber a by high-power laser can be used to adjust the ratio between the size and doping concentration of the microspheres. A doped fiber was tapered and cut as before. However, this time no undoped fiber section was used. After cutting with the laser, the remaining fiber length was recorded as \emph{l${_d}$}. The CO$_{2}$ laser was set to 7 W and kept on as the fiber was slowly moved downward and fed into the focus of the laser, during which the glass melted and evaporated into a microsphere. When the tapered region was reached, the sphere was held in the focus until it evaporated down to the desired size.

The mols of Er$^{3+}$ lost (\emph{n${_{Er loss}}$}) can be calculated from the mols of SiO$_{2}$ lost when evaporating the volume of the fiber down to the volume of the sphere:
\begin{equation}
\emph{n${_{Er loss}}$} =\emph{R${_{mm}}$}\times
 \frac{\rho[(\pi\times\emph{r${_{f}}$}^2)\emph{l${_d}$}-(4/3)(\pi\times\emph{r${_{s}}$}^3)]}{\rm M}
\end{equation}
The molar amount of Er$^{3+}$ in the core (\emph{n${_{Er core}}$}) is:
\begin{equation}
\emph{n${_{Er core}}$} =\emph{p${_{mpc}}$}\times
 \frac{\rho(\pi\times\emph{r${_{c}}$}^2)\emph{l${_d}$}}{\rm M}
\end{equation}
The molar amount of SiO$_{2}$ in the microspheres (\emph{n${_{SiO2 sphere}}$}) is:
\begin{equation}
\emph{n${_{SiO2 sphere}}$} =
 \frac{\rho(4/3)(\pi\times\emph{r${_{s}}$}^3)}{\rm{M}}
\end{equation}
The molar percentage of Er$^{3+}$ in the doped microspheres (\emph{p${_{mps}}$}) formed by the evaporation of the doped fiber can be calculated:
\begin{equation}
\emph{p${_{mps}}$} =
 \frac{\emph{n${_{Er core}}$}-\emph{n${_{Er loss}}$}}{\emph{n${_{Er core}}$}-\emph{n${_{Er loss}}$}+\emph{n${_{SiO2 sphere}}$}}
\end{equation}

\begin{figure}[htbp]
\centering\includegraphics[width=18cm]{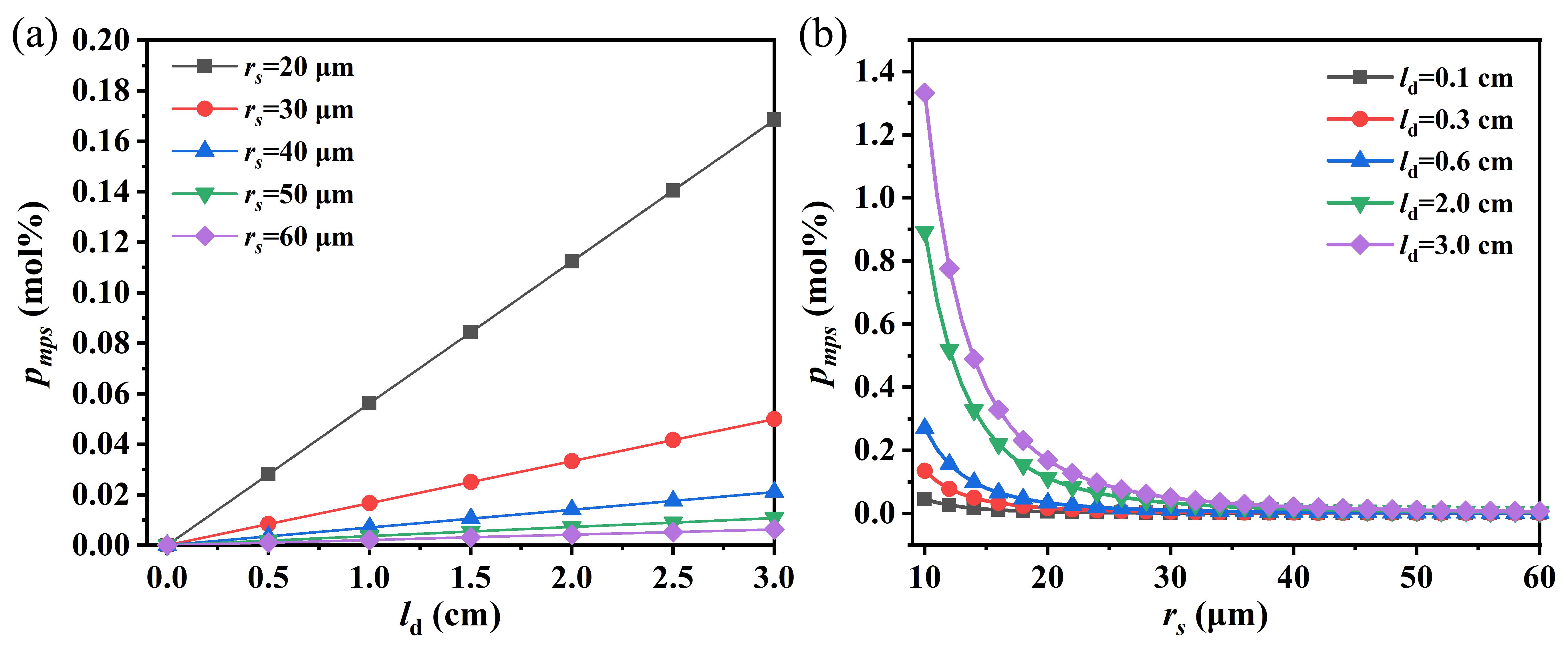}
\caption{Molar percentage concentration of microspheres with different parameters. (a) Effect of doped fiber length on concentration for radii ranging from 20 $\mu$m to 60 $\mu$m. (b) Effect of microsphere radius on concentration for fiber lengths ranging from 0.1 cm to 3 cm.}
\label{fig equation 8}
\end{figure}

From equation (8), one can see that for a given sphere size, the doping concentration scales linearly with the \emph{l${_d}$} but for a fixed \emph{l${_d}$} the concentration scales nonlinearly with sphere size, as shown in Figure \ref{fig equation 8}(a) and (b). Therefore, one needs to consider the length of the fiber to be evaporated for a given sphere size to arrive at the final concentration. For example, it should be possible to increase the final concentration of the sphere to three or four times higher than the original concentration of the doped fiber by choosing the appropriate length to be evaporated. For example, when a 3 cm length of  fiber, as used in these experiments, was evaporated down to a sphere with a 20 $\mu$m radius, the volume was reduced by a factor of 11,000 and inc the concentration was increased by a factor of 3.4 from 0.05$\%$ to 0.17$\%$. The fiber used here was the same as that in Figure \ref{lu_and_ld}, hence \emph{L${_{mv}}$}=$6.36 \times$ 10$^{-21}$ mol/$\mu$m$^{3}$ and \emph{R${_{mm}}$}=$1.74 \times$ 10$^{-7}$.

\begin{figure}[h]
\centering\includegraphics[width=18cm]{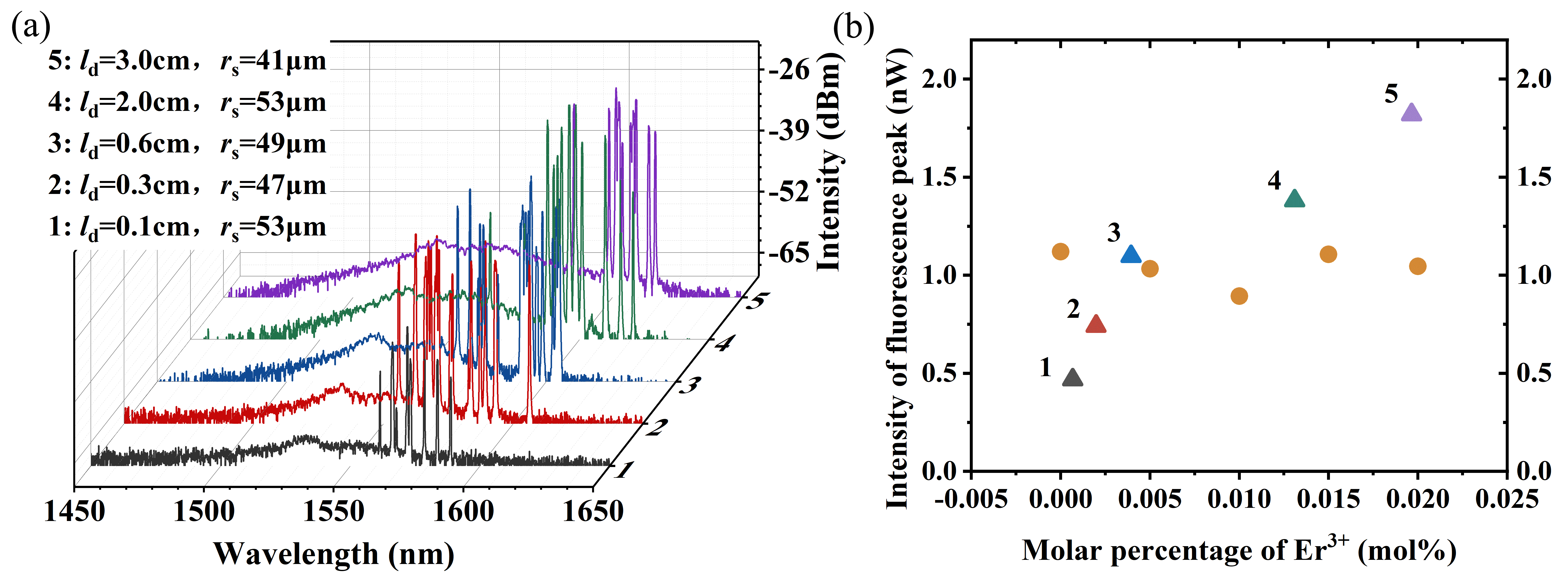}
\caption{(a) Spectra of five samples prepared by evaporation under the same excitation conditions. (b) The peak fluorescence intensity as a function of the calculated Er$^{3+}$ concentration for the five different samples (triangles) and the variation of the peak fluorescence intensity of Sample three for five different coupling positions (dots).}
\label{samples}
\end{figure}

Five microspheres were prepared directly from doped fiber of different lengths of \emph{l${_d}$}, while the radius of the spheres was around 50 $\mu$m. Figure \ref{samples}(a) shows the spectrum of each sample under the same excitation conditions. The fluorescent bandwidth and power are, respectively, wider and stronger for the microspheres made from longer lengths of \emph{l${_d}$} as a result of higher concentrations. Figure \ref{samples}(b) shows the fluorescence peak power plotted against the calculated Er$^{3+}$ concentration, which appears to be positively correlated with the fluorescence intensity. These experimental results confirm the ability of the preparation method to control the concentration, hence, the intensity of the fluorescence.

Additionally the laser emission tends to longer wavelength, possibly due to the reduced distance between Er$^{3+}$ enhancing the inter-ion energy transfer and excited state absorption \cite{doi:10.1021/cm0301930}. The absorption spectrum of Er$^{3+}$ shows that, for the light emitted by Er$^{3+}$ in the 1550 nm band, the shorter wavelength is easier absorbed by nearby Er$^{3+}$ \cite{YAMAUCHI2005679}, so the absorption from adjacent Er$^{3+}$ enhances laser emission at long wavelengths. 

Because lasing peaks in the spectra are extremely sensitive to the coupling condition with the tapered fiber, the relationship between laser intensity and Er$^{3+}$ concentration in microspheres is not obvious. As can be seen from the spectra, unlike the high sensitivity of the lasing peaks to variations in the coupling, the fluorescence intensity is less affected by the coupling condition. The orange dots in Figure \ref{samples}(b) represent the fluorescence peak power of Sample 3 at different coupling positions. When the lasing peaks are effectively excited, the coupling position has little effect on the fluorescence intensity.  

\section{Conclusion}

In conclusion, the loss rate of Er$^{3+}$ in doped silica fiber during evaporation was estimated by studying the fluorescence emission from WGRs. By controlling the volume of the evaporated glass fiber and determining the glass volume where the fluorescence emission disappeared, it was found that Er$^{3+}$ was lost in the evaporation process at a rate of $6.36 \times$ 10$^{-21}$ mol of Er$^{3+}$ for every cubic micron of silica fiber. This can also be expressed as ratio of $1.74 \times$ 10$^{-7}$ mol of Er$^{3+}$ lost for each mol of SiO$_{2}$ evaporated. 

The result demonstrates a method to estimate the evaporation in the low doping concentration range. In principle, this technique could be used to study other rare earth ions and luminescent materials where the relevant data is helpful in the preparation and characterisation of doped glass photonic devices. In summary, active microsphere resonators were directly prepared from doped fiber such that the final erbium concentration in the microsphere could be increased or decreased relative to the initial concentration in the fiber. This is a study and demonstration of controlled doping concentration in whispering gallery resonators that does not require chemical processing such as sol-gel or the fabrication of specifically doped fiber to achieve the desired concentration.

% Acknowledgements
\medskip
\textbf{Acknowledgements} \par %delete if not applicable))
National Natural Science Foundation of China (62225502, 61935006, 62090062); Okinawa Institute of Science and Technology Graduate University; CAS Interdisciplinary Innovation Team project (JCTD-2018-19).

% References
\medskip

% Use the following code if you wish to generate your bibliography with BibTeX;
% replace the string "MSP-template" below with the name(s) of
% the BibTeX data base(s) you want to use.
% The resulting bibliography-output (the content of the .bbl file)
% must be pasted back into this file before submission.
% Please also include your BibTeX data base file(s) in your submission
% so that we can re-run BibTeX if necessary.
%
\bibliographystyle{MSP}
\bibliography{Ref}

\begin{table}[htbp]
	\centering
	\caption{The data of 5 samples.}
	\label{tab:1}
	\resizebox{\textwidth}{!}{
	\begin{tabular}{cccccc}
		\noalign{\smallskip}\hline
		Sample & 1 & 2 & 3 & 4 & 5  \\
		\noalign{\smallskip}\hline
		\emph{l${_d}$} (cm) & 0.1 & 0.3 & 0.6 & 2.0 & 3.0  \\
		\emph{r${_{s}}$} ($\mu$m) & 53 & 47 & 49 & 55 & 41  \\
		Molar percentage of Er$^{3+}$ (mol$\%$) & 0.00032 & 0.00132 & 0.00231 & 0.00543 & 0.01961  \\
		Intensity of fluorescence peak (dBm) & -63.3 & -61.3 & 59.6 & -58.6 & 57.4  \\
		\noalign{\smallskip}\hline
	\end{tabular}}
\end{table}

\end{document}